\documentclass{article}

\usepackage{PRIMEarxiv}

\usepackage[utf8]{inputenc} 
\usepackage[T1]{fontenc}    
\usepackage{hyperref}       
\usepackage{url}            
\usepackage{booktabs}       
\usepackage{amsfonts}       
\usepackage{nicefrac}       
\usepackage{microtype}      
\usepackage{lipsum}
\usepackage{fancyhdr}       
\usepackage{graphicx}       
\usepackage{amsmath}
\usepackage{quantikz}
\graphicspath{{media/}}     

\title{GPU benchmark through QPE emulator with cuQuantum for practical quantum applications
}

\author{
  Takaki AKIBA \\
  Department of Mechanical Engineering\\
  The University of Tokyo \\
  Tokyo, Japan\\
  \texttt{akiba@mesl.t.u-tokyo.ac.jp} \\
   \And
  Youhi MORII \\
  Institute of Fluid Science \\
  Tohoku University \\
  Sendai, Japan\\
  \texttt{morii@edyn.ifs.tohoku.ac.jp} \\
}

\begin{document}
\maketitle

\begin{abstract}
The quantum algorithm of Quantum Phase Estimation (QPE) was implemented to make the maximum use of GPU emulation with cuQuantum and CUDA Toolkit by NVIDIA. 
The input and output were handled by HDF5 to make the process as easy as possible. 
The computational time, VRAM usage, value error, and overhead was evaluated against the developed application.
VRAM usage and the profiler analysis suggested that the developed application could make the maximum use of GPU capability.
\end{abstract}


\section{Introduction}
Quantum algorithms are designed to leverage the capabilities of quantum computers 
and many have been proposed in recent years.
Thanks to the efforts of many developers, various frameworks are available which enable the execution of these algorithm on quantum emulators or
small-scale quantum devices, often without financial cost \cite{javadi-abhari_quantum_2024, developers_cirq_2025, bergholm_pennylane_2022}.
However, the examples provided by such frameworks are typically limited to very small systems -up-to 10 qubits- which can be easily simulated on classical computers.
While some studies have reported the use of GPUs for mid-scale quantum emulation, currently available sample codes are scarce.
From the perspective of computer-aided engineering (CAE), 
there is increasing demand for practical-scale quantum emulators that can provide hands-on experience with quantum computation.
This raises several important questions:
\begin{itemize}
  \item What types of problems can be addressed using quantum computers?
  \item What forms of output can be expected after quantum processes?
  \item What data representations are most suitable for quantum computing?
\end{itemize}
In this work, we aim to provide a concrete example of implementing a quantum algorithm on GPUs, 
scaling up to the limit imposed by GPU hardware, mainly by GPU VRAM capacity.
We focused on Quantum Phase Estimation (QPE) \cite{kitaev_quantum_1995, ahmadi_quantum_2011}, 
a representative and fundamental quantum algorithm which is also used as a part of many other quantum algorithms \cite{hallgren_polynomial-time_2007, shor_algorithms_1994, shor_polynomial-time_1997, szegedy_quantum_2004, Harrow2009}.

\section{Quantum Phase Estimation}
Quantum Phase Estimation (QPE) is a fundamental quantum algorithm used to determine the eigenvalues of a given unitary matrix.
Specifically, QPE estimates the phase $\phi_i$ in the eigenvalue expression
$$
\lambda_i = \exp(2\pi i \phi_i), 0 \le \phi_i  \le 1,
$$
where $\lambda_i$ is the $i$-th eigenvalue of the unitary matrix and $\phi_i$ is referred to as the eigenphase.
The QPE algorithm requires the following conditions for its implementation:
\begin{itemize}
\item The unitary matrix must be decomposable into a quantum circuit, including controlled-unitary operations.
\item The corresponding eigenvector must be provided and initialized on the quantum circuit as the input statevector.
\end{itemize}
The quantum circuit for QPE is illustrated in Figure \ref{fig:figqpe}.
It consists of two groups of quantum registers: one for phase estimation (measurement qubits) 
and the other for applying the unitary operator (target qubits).
QPE introduces two key parameters that independently influence both computational cost and estimation accuracy: 
\begin{itemize}
    \item $N_{meas}$ :the number of measurement qubits
    \item $N_{mat}$ :the number of qubits representing the target matrix and the eigenvector.
\end{itemize}
Understanding how these parameters affect performance and accuracy is a central objective of this work.

\def\myvdots{\ \vdots\ }
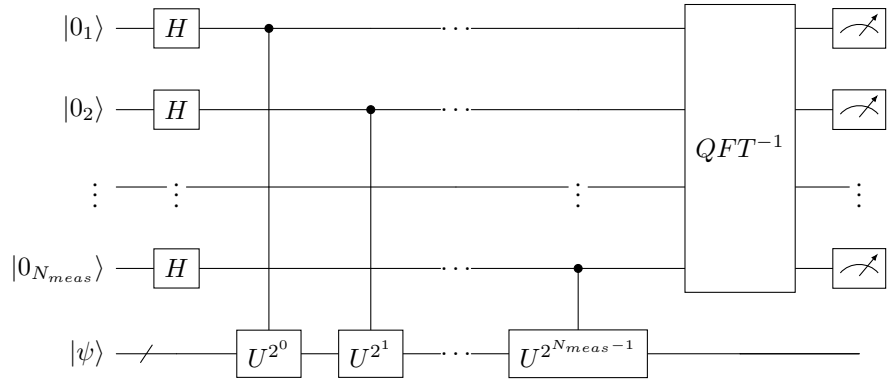
\begin{figure}
  \centering
  \begin{quantikz}[thin lines]
    \lstick{{$\ket{0_1}$}}          & \gate{H}    & \ctrl{4}         &                  & \ldots &                           & \gate[wires=4]{QFT^{-1}} & \meter{} \\
    \lstick{{$\ket{0_2}$}}          & \gate{H}    &                  & \ctrl{3}         & \ldots &                           &                          & \meter{} \\
    \lstick{\myvdots}               & \myvdots    &                  &                  &        & \myvdots                  &                          & \myvdots \\
    \lstick{{$\ket{0_{N_{meas}}}$}} & \gate{H}    &                  &                  & \ldots & \ctrl{1}                  &                          & \meter{} \\
    \lstick{$\ket{\psi}$}           & \qwbundle{} & \gate{U^{2^{0}}} & \gate{U^{2^{1}}} & \ldots & \gate{U^{2^{N_{meas}-1}}} &                          & \qw \\
  \end{quantikz}
  \caption{Quantum circuit of QPE}
  \label{fig:figqpe}
\end{figure}

\section{Concept design}
\label{sec:headings}

\begin{figure}
  \centering
  \includegraphics[width=\columnwidth]{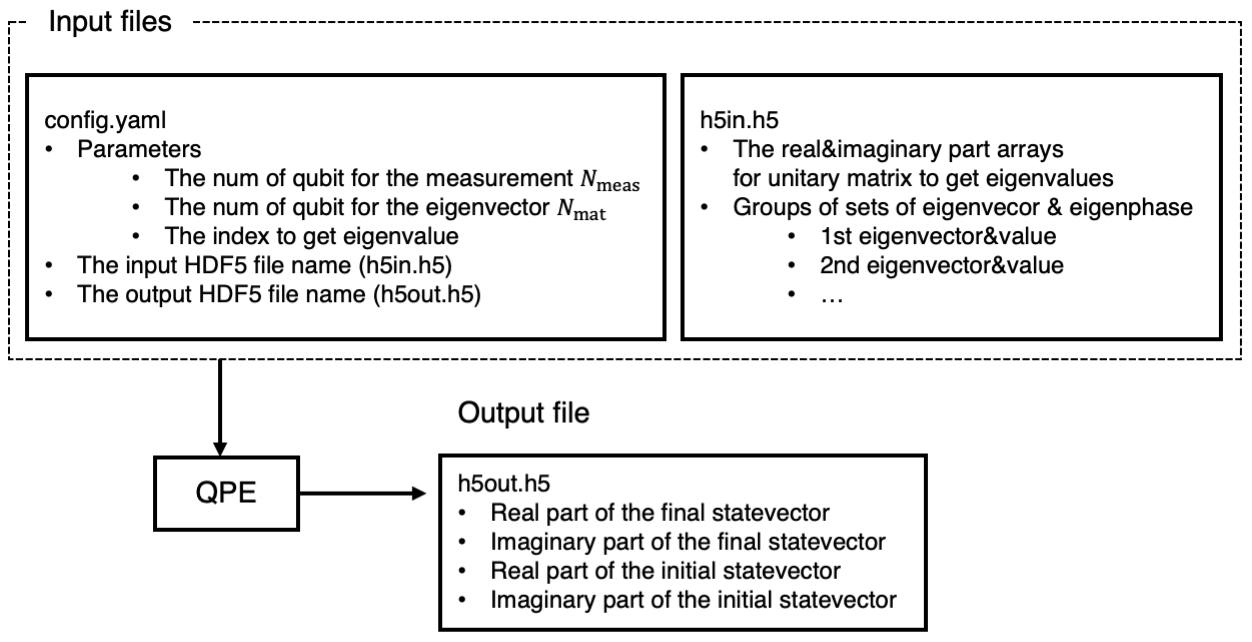}
  \caption{Structure and interfaces of the presented application}
  \label{fig:figc}
\end{figure}

The application structure and interfaces are presented in Figure \ref{fig:figc}.
To ensure general usability, the implementation follows several key design principles.
The target unitary matrix and its corresponding eigenvector are supplied via HDF5 \cite{the_hdf_group_hierarchical_nodate} files,
which facilitate both pre-processing and post-processes.
The matrix and eigenvector dimensions must strictly be $ 2^{N_{mat}} \times 2^{N_{mat}} $ and $ 2^{N_{mat}} $, respectively,
where $ N_{mat} $ denotes the number of qubits used to represent the target matrix.

The main emulation program is implemented using NVIDIA's cuQuantum library \cite{bayraktar_cuquantum_2023} and the CUDA toolkit. 
It reads the target matrix and eigenvector from the HDF5 input file, and computes the corresponding eighenphase using the QPE algorithm.
As quantum simulators allow access to the full statevector, the final statevector is written to another HDF5 file for the post-processing. 
The estimated eigenphase can then be extracted by analyzing the norm of the final statevector.

Two types of test problems were used to validate the implementation. 
The first is an analytically trackable case defined as follows: 

\begin{gather}
 \phi_j = \frac{j}{ 2^{ N_{mat} } } \\
 \bold{x_j} = \frac{1}{\sqrt{j}}\{ x_j^k \},  \\
 x_j^k = 
 \begin{dcases*}
     1 & if $k \le j$ , \\
     0 & otherwise
 \end{dcases*}
\end{gather}
 
Using these eigenphases and eigenvectors, the target unitary matrix $ U $ is constructed by:

\begin{gather}
    \Lambda = diag\left(
    \exp(2 \pi i \phi_0 ),
    \exp(2 \pi i \phi_1 ),
    \cdots,
    \exp(2 \pi i \phi_{N_{mat}-1} )
    \right) \\
    P = \left( 
    \bold{x}_0, \bold{x}_1, \cdots, \bold{x_{N_{mat}-1}}
    \right) \\
    U = P \Lambda P^{-1}
\end{gather}
 
As the eigenphases are encoded using $ 2^{N_{meas}} $ resolution, they can, in theory, be estimated with zero error. 
This test case therefore serves as a debugging and validation benchmark for the implementation. 

The second test case involves random unitary matrices, 
generated using the Qiskit library \cite{javadi-abhari_quantum_2024} with a fixed random seed (12345). 
Reference eigenphases and eigenvectors are separately computed using SciPy \cite{2020SciPy-NMeth} linear algebra (scipy.linalg module.

The QPE simulations were executed on various NVIDIA GPUs, including of GeForce RTX 4080 Super, A100, and GH100. 
The performance profiling was conducted using the NVIDIA Profiler after implementation.  

\section{Validations}
\label{sec:others}
This section presents validation results for the developed QPE implementation, focusing on three aspects:
computational time, estimation accuracy, and runtime profiling on GPUs.

\subsection{Computational time}
The computational time required for QPE execution was evaluated as a function of the total number of qubits used in the simulator. 
To isolate performance effects from numerical errors,
only test matrices with analytically known eigenphases were used in this analysis.

\begin{figure}
  \centering
  \begin{minipage}{0.43\columnwidth}
    \centering
    \includegraphics[width=\columnwidth]{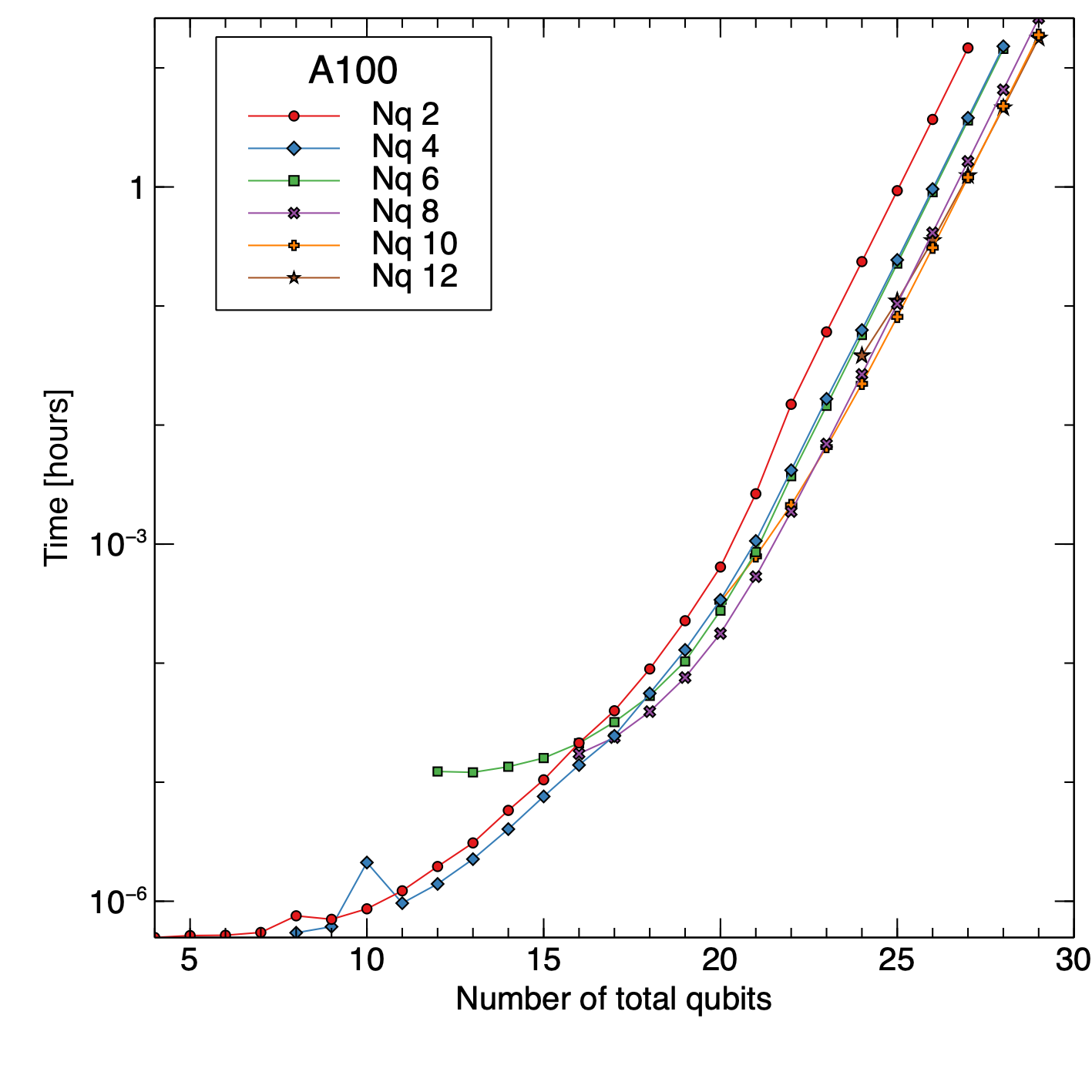}
    \caption{Computational time vs number of total qubits in QPE on A100 GPU}
    \label{fig:fig1}
  \end{minipage}
  \hspace{5mm}
  \begin{minipage}{0.43\columnwidth}
    \centering
    \includegraphics[width=\columnwidth]{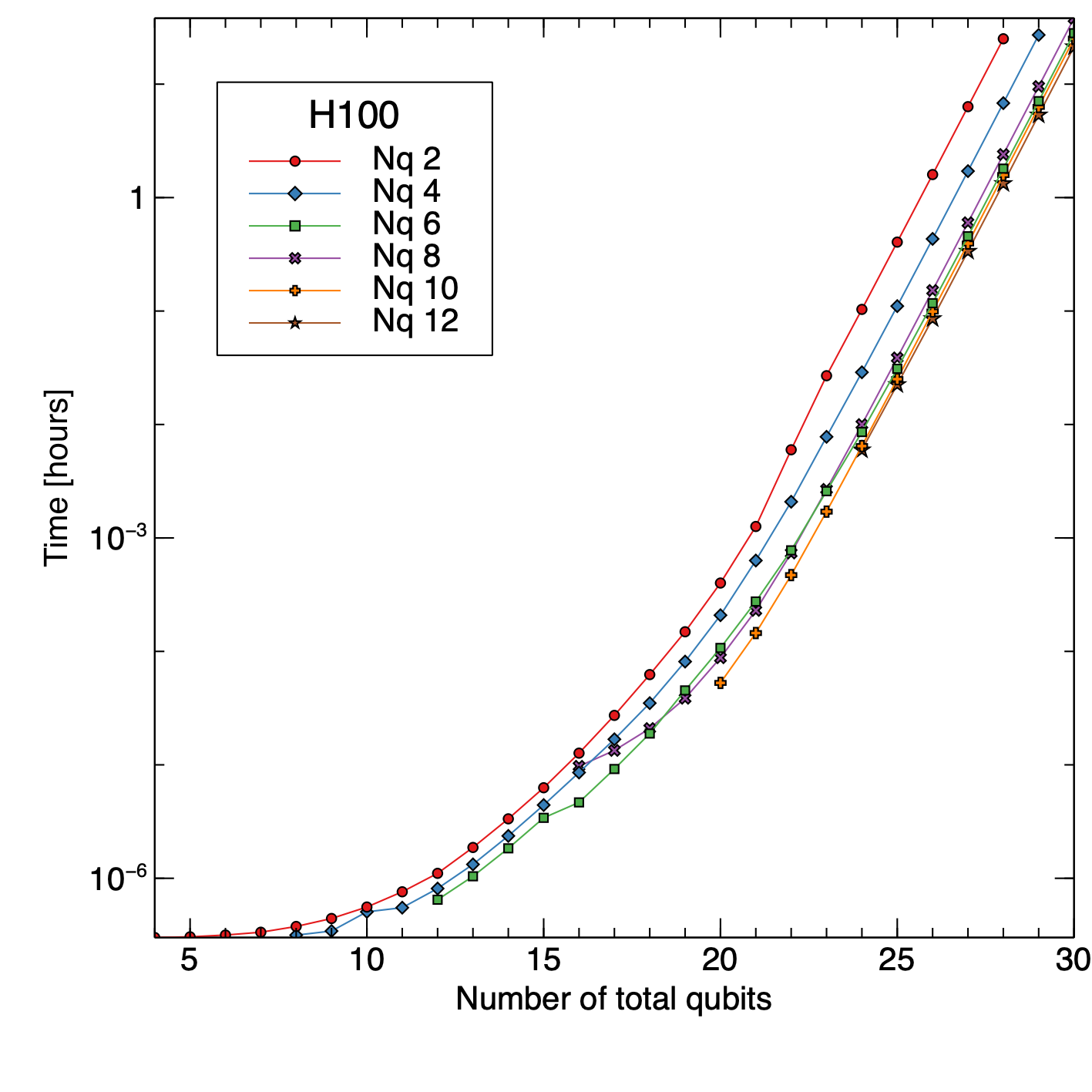}
    \caption{Computational time vs number of total qubits in QPE on H100 GPU}
    \label{fig:fig2}
  \end{minipage}

  \vspace{3mm}
  
  \begin{minipage}{0.43\columnwidth}
    \centering
    \includegraphics[width=\columnwidth]{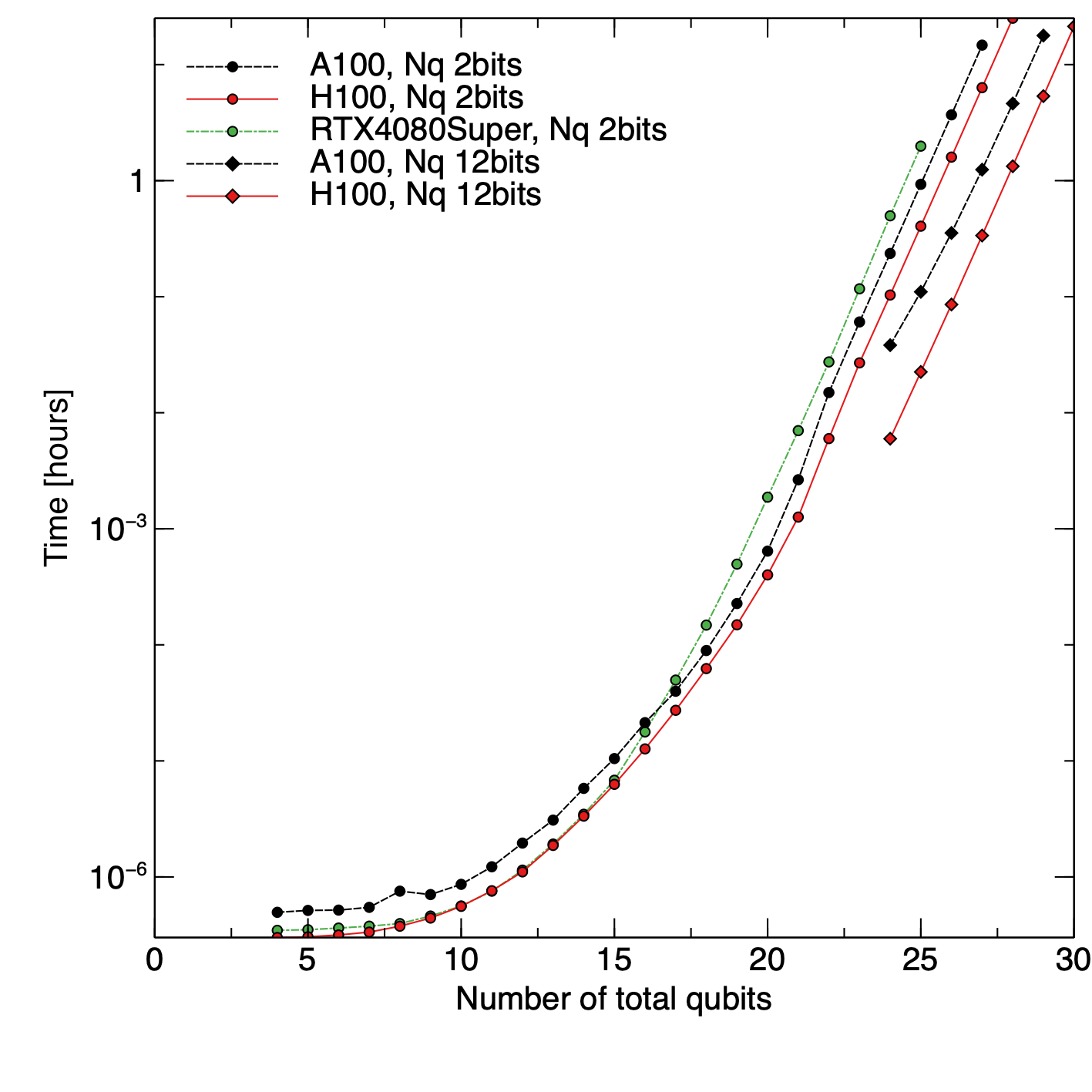}
    \caption{Comparison of computational time of QPE on different GPUs}
    \label{fig:fig3}
  \end{minipage}
  \hspace{5mm}
  \begin{minipage}{0.43\columnwidth}
    \centering
    \includegraphics[width=\columnwidth]{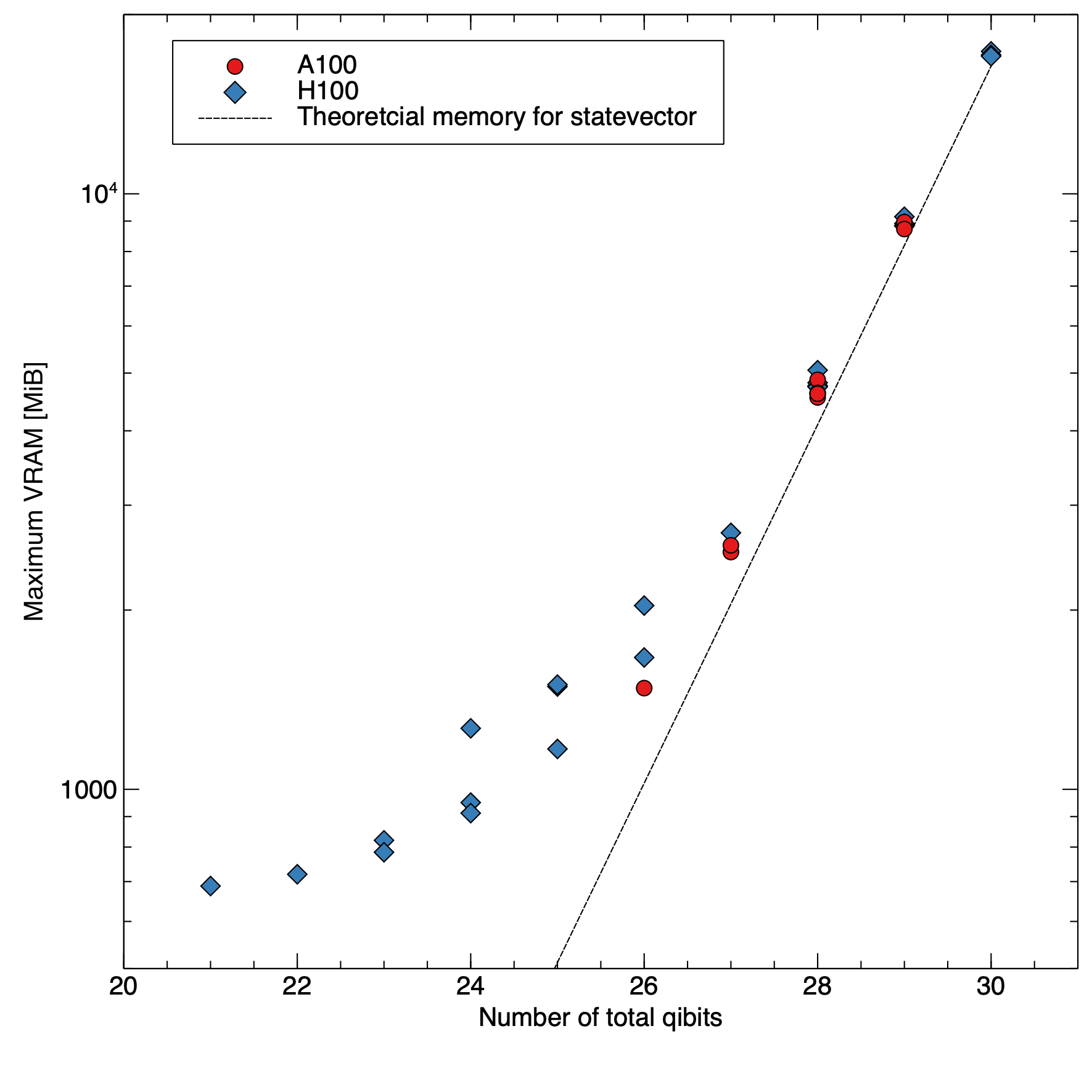}
    \caption{GPU memory usage against the total number of qubits}
    \label{fig:fig4}
  \end{minipage}
  
\end{figure}

Figure \ref{fig:fig1} shows the computational time (in hours) versus the total number of qubits.
Here, the total number of qubits $N_{total}$ is defined as the sum of qubits used for measurement and for representing the target unitary matrix: 

\begin{gather}
  N_{total} = N_{meas} + N_{mat}.
\end{gather}

At higher values of $N_{total}$, the computational costs increases approximately with the fourth power of $N_{total}$. 
This scaling behavior is attributed to two main factors: 
\begin{enumerate}
    \item The number of controlled-unitary operation grows exponentially (by a factor of 2) with the number of measurement qubits.
    \item Each application of the unitary operator becomes more computationally intensive as the statevector size increases, also doubling with each additional qubit.
\end{enumerate}
Taken together, these effects result in a quadratic relationship between computational cost and $N_{total}$ in log-log scale. 
At smaller qubit counts, however, the computational time shows signs of saturation, which can be attributed to overheads such as CUDA kernel launch latency and CPU-GPU communication. 
These aspects will be discussed in more detail in the profiling analysis section.

Similar scaling behavior was observed on the NVIDIA H100 GPU, as shown in Figure \ref{fig:fig2}.

Figure \ref{fig:fig3} compares performance across different GPU.
The observed computational times reflect the relative performance of the hardware: 
the GeForce RTX 4080 Super exhibited the slowest runtime, while the GH100 achieved the fastest.

Figure \ref{fig:fig4} shows the peak GPU memory usage as a function of the total number of qubits.
The dotted line shows the theoretical minimum memory required to store the full statevector in double precision.
At small qubit counts, actual memory usage exceeds this theoretical bound due to the additional memory required to implement certain quantum gates (e.g., Hadmard gates) in simulation. 
While such overhead is significant in small-scale simulations, its relative impact diminishes as the statevector size increases.

\subsection{Evaluation of precision against random unitary problem}
In this analysis, the matrix size was fixed at 16x16, corresponding to a 4 qubit system.
The accuracy of the estimated eigenphase was evaluated using two different methods in order to assess the impact of the evaluation strategy. 
The first method selects the single basis state with the maximum norm.
This approach is practical, especially in real quantum systems where measurement outcomes are noisy and only the most probable state may be reliably observed.

The second method computes a weighted average of all possible measured eigenphases, with weights proportional to the squared norms of the corresponding amplitudes. 

For each method, the estimated eigenphases were compared with reference values computed using the scipy.linalg module. 
The individual errors were calculated as the absolute differences between the estimated and reference eigenphases. 
These were then averaged over all 16 eigenphases to yield a representative error metric.

This evaluation was repeated while varying the number of measurement qubits used in the QPE circuit. 
The goal was to investigate how the number of measurement qubits influences estimation accuracy. 
The results are presented in Figure 6, which shows the representative error as a function of measurement qubit count.

As theoretically expected, the estimation accuracy improves exponentially with the number of measurement qubits, roughly halving the error for each additional qubit. 
Regarding the comparison between evaluation methods, the maximum-norm method yields better accuracy when the number of measurement qubits is relatively small. 
However, beyond a certain threshold (denoted as $N_{meas} = 21$), the weighted-average method outperforms, owing to its ability to incorporate broader amplitude information.

\begin{figure}
  \centering
  \includegraphics[width=\columnwidth]{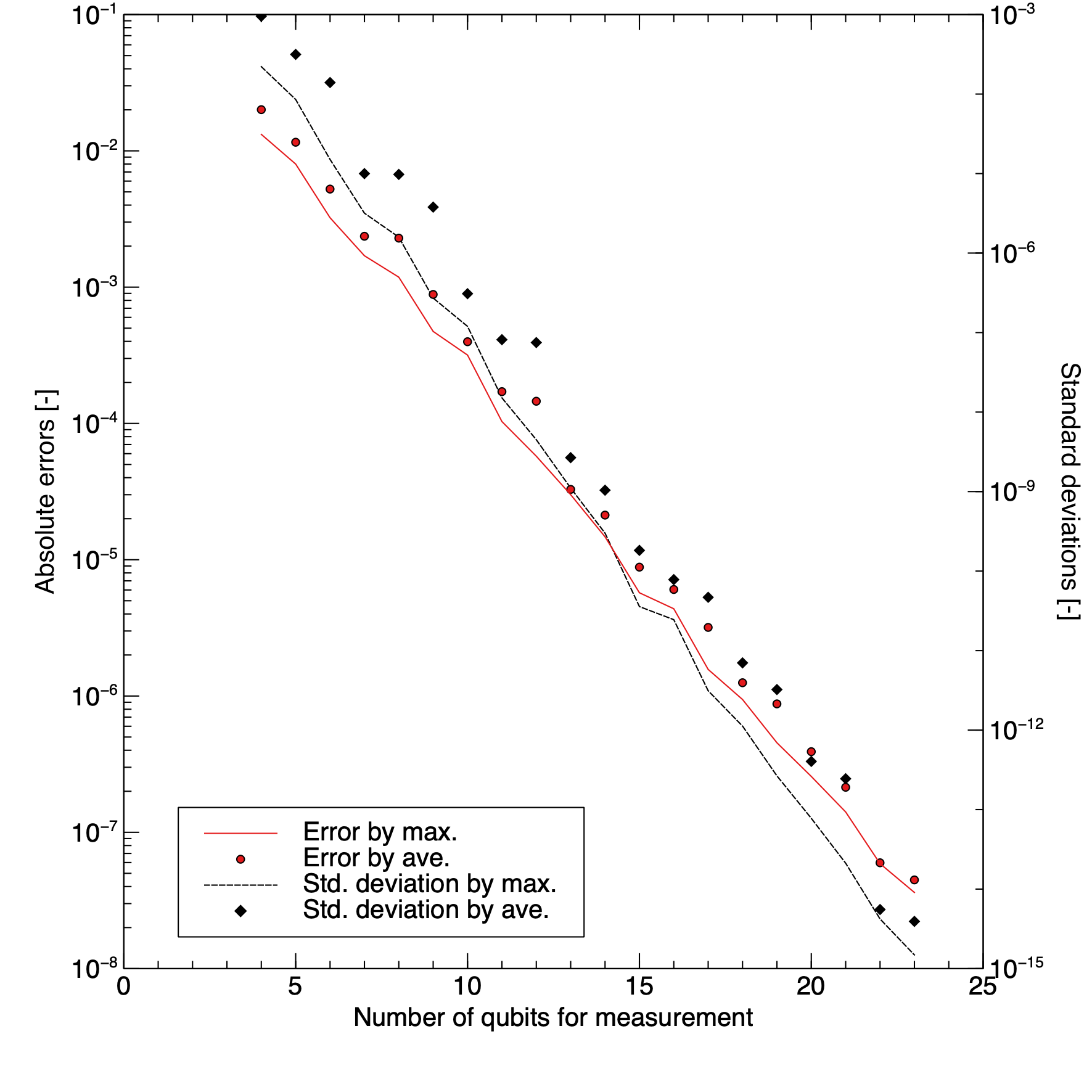}
  \caption{Numerical precision vs number of qubits for measurement bit}
  \label{fig:fig5}
\end{figure}

\subsection{Profiler analysis with NVIDIA Profiler}
The application was analyzed using the NVIDIA Profiler to identify performance bottlenecks. 
Three dominant processes were identified:
\begin{itemize}
    \item GPU memory allocation
    \item CPU-to-GPU memory transfer
    \item Application of the unitary matrix
\end{itemize}
The GPU memory allocation remained almost constant at approximately 110-120 ms, regardless of the number of qubits.

The memory copy process exhibited two distinct regimes. 
As shown in Table \ref{tab:table1}, when $N_{mat} \le 5$, the elapsed time was consistently less than 5 us and showed little dependence on $N_{meas}$.
However, for $N_{mat} > 5$, the elapsed time increased significantly and approximately doubled with each additional measurement qubit.

Table \ref{tab:table2} summarizes the results for the matrix application process.
At a fixed $N_{mat}$, the elapsed time increased significantly and approximately doubled with each increment in $N_{meas}$.

\begin{table}
  \centering
  \caption{The effect of qubits for the target unitary matrix and measurement on the memory copy process.}
  \begin{tabular}{rrr}
    \hline
    $N_{mat}$ & $N_{meas}$ & elapsed time for memory copy [us] \\
    \hline
    4         & 15         & 1.8 \\
    4         & 14         & 1.8 \\
    5         & 15         & 2.4 \\
    5         & 16         & 3   \\
    6         & 14         & 390 \\
    6         & 15         & 730 \\
    8         & 12         & 1550 \\
    8         & 13         & 2850 \\
    \hline
  \end{tabular}
  \label{tab:table1}
\end{table}

\begin{table}
  \centering
  \caption{The effect of qubits for the target unitary matrix and measurement on matrix application process.}
  \begin{tabular}{rrr}
    \hline
    $N_{mat}$ & $N_{meas}$ & elapsed time for memory copy [us] \\
    \hline
    2         & 18         & 25 \\
    2         & 19         & 47 \\
    4         & 14         & 60 \\
    4         & 15         & 107 \\
    4         & 16         & 190 \\
    5         & 15         & 210 \\
    5         & 16         & 370 \\
    \hline
  \end{tabular}
  \label{tab:table2}
\end{table}
  
\section{Conclusion}
In this study, we implemented the Quantum Phase Estimation algorithm using the cuQuantum library and CUDA to fully utilize the computational power of GPUs.
The implementation was designed to scale up to the limits of GPU VRAM capacity, 
and was evaluated in terms of computational cost, memory usage, estimation accuracy, and runtime overhead.
Two types of test problems were introduced: one with analytically defined eigenphases that can be expressed exactly in binary form,
and another using randomly generated unitary matrices with general eighenphases.
The computational cost was found to scale approximately with the fourth power of the total number of qubits.
Performance comparisons across different GPUs confirmed that higher-end hardware yielded faster execution times.
Estimation accuracy improved exponentially with the number of measurement qubits, 
validating theoretical expectations.
Additionally, profiling revealed that memory allocation and CPU-GPU data transfer can become dominant overheads, 
especially as system size increased.
These result demonstrate that the proposed implementation can server as a practical foundation for mid-scale quantum algorithm emulation and provide valuable insights for quantum computing applications in engineering contexts.

\section{Source code}
The QPE application with GPU emulation used in this study is available at https://github.com/takakiba/CuQuEmBench.git.

\section*{Acknowledgments}
This was partially supported by JSPS Grant-in-Aid for Research Activity Startup (24K22922), 
Supercomputer Center, the Institute of Solid State Physics, The University of Tokyo, 
and Recommendation Program for Young Researchers and Woman Researchers, Supercomputing Division, Information Technology Center, 
The University of Tokyo.

\bibliographystyle{unsrt}  
\bibliography{cuda_qpe}

\end{document}